\documentclass[onecolumn,aps,superscriptaddress]{revtex4-1}

\usepackage{amsmath, amssymb}
\usepackage{graphicx}

\begin{document}


\title{Intertwined vestigial order in quantum materials: nematicity and beyond}

\author{Rafael M. Fernandes}
\affiliation{School of Physics and Astronomy, University of Minnesota, Minneapolis,
Minnesota 55455, USA}
\author{Peter P. Orth}
\affiliation{Department of Physics and Astronomy, Iowa State University, Ames,
Iowa 50010, USA}
\author{J\"org Schmalian}
\affiliation{Institute for Theory of Condensed Matter and Institute for Solid
State Physics, Karlsruhe Institute of Technology (KIT), Karlsruhe,
Germany}

\begin{abstract}
A hallmark of the phase diagrams of quantum materials is the existence of multiple electronic ordered states, which, in many cases, are not independent competing phases, but instead display a complex intertwinement. In this review, we focus on a particular realization of intertwined orders: a primary phase characterized by a multi-component order parameter and a fluctuation-driven vestigial phase characterized by a composite order parameter. This concept has been widely employed to elucidate nematicity in iron-based and cuprate superconductors. Here we present a group-theoretical framework that extends this notion to a variety of phases, providing a classification of vestigial orders of unconventional superconductors and density-waves. Electronic states with scalar and vector chiral order, spin-nematic order, Ising-nematic order, time-reversal symmetry-breaking order, and algebraic vestigial order emerge from one underlying principle. The formalism provides a framework to understand the complexity of quantum materials based on symmetry, largely without resorting to microscopic models.
\end{abstract}

\maketitle


\section{INTRODUCTION}
Many quantum materials are characterized by a rich phase diagram,
in which numerous order parameters assume finite values in neighboring
regions of the parameter space of temperature, chemical composition,
mechanical strain, and electromagnetic fields. The natural instinct
one has to capture this physics is to assign the different phases
to competing order parameters. For example, if one finds antiferromagnetism
(with order parameter $\mathbf{m}$) and superconductivity (with order
parameter $\Delta$) nearby, one writes individual Ginzburg-Landau
expansions for the free energy $f_{m}$ and $f_{\Delta}$ for both
degrees of freedoms, coupled by a symmetry-allowed term such as 
$f_{m-\Delta}=\gamma\mathbf{m}\cdot\mathbf{m}\left|\Delta\right|^{2}$.  
Positive $\gamma$ amounts to phase competition while negative $\gamma$
causes one phase to attract the other. While this approach proved
to be very efficient in many cases \cite{Sachdev2004,Fernandes2010, Fernandes2010-2},
Landau theory cannot explain why multiple phases emerge close to each
other in a phase diagram. Addressing this question usually requires
a microscopic description in terms of a model Hamiltonian, a task
that can be technically challenging. Given the abundance of complex
phase diagrams in correlated electronic systems, it is desirable to
identify general principles to describe the close relationship between
their multiple ordered states.

An underlying general principle to rationalize complex phase diagrams
without necessarily resorting to a microscopic description was recently
advocated in Ref. \cite{Fradkin2015} and is generally referred to
as intertwined order. The idea is that multiple phases of a rich phase
diagram are born out of a primary state. A prime example for such
a behavior is that of pair-density-wave order, which entangles superconductivity
and density waves \cite{Himeda2002,Agterberg08,Berg09,Berg2009, Berg2009-2,Loder2011}.
Intertwined orders can also arise due to the interactions induced
by a primary order parameter near a quantum phase transition. For
example, antiferromagnetic or nematic fluctuations near quantum critical
points have been proposed to provide or enhance the pairing interactions
for a superconducting phase \cite{Chubukov2003,Lederer2008}. 

In this review, we focus on a particular realization of intertwined
phases in terms of vestigial \textendash{} or composite \textendash{}
order. Composite order exists when higher order combinations of potentially
symmetry-breaking order parameters condense. Consider a complex multi-component
field $\eta_{\alpha}$, where $\alpha$ labels the order parameter
components. A finite expectation value $\left\langle \eta_{\alpha}\right\rangle $
would break a certain symmetry of the system \textendash{} for instance,
time-reversal in the case of ferromagnetism or translational symmetry
in the case of charge order. Composite order then corresponds to the
case where certain combinations of the product of the order parameters
are on average non-zero, whereas each individual order parameter remains
zero on average:
\begin{equation}
\left\langle \eta_{\alpha}^{*}\eta_{\beta}\right\rangle \neq0\qquad\mathrm{but}\qquad\left\langle \eta_{\alpha}\right\rangle =0.\label{eq:composite intro}
\end{equation}

The bilinear combination $\left\langle \eta_{\alpha}^{*}\eta_{\beta}\right\rangle $
behaves itself as an order parameter, which breaks only a subset of
the symmetries broken by $\eta_{\alpha}$. For this reason, the composite
order is called a \emph{vestige} of the primary phase where $\left\langle \eta_{\alpha}\right\rangle $
is finite. This makes both the composite and primary orders naturally
intertwined. At first glance, this scenario may seem rather contrived.
However, as we will show here, it naturally arises in many quantum
materials, when the primary order parameter has multiple components,
such that the primary phase is degenerate. In Eq. (\ref{eq:composite intro})
we allowed for $\eta_{\alpha}$ to be complex. This is relevant for
superconductors or incommensurate density-wave states. 

There are two complementary ways to approach a composite ordered phase.
If one starts from the primary ordered phase, composite order can
be understood as a partial melting of the former, before the system
goes to a completely disordered phase \cite{Kivelson98,Zaanen04}.
Conversely, starting from the disordered phase, vestigial order can
be understood as a fluctuation-induced composite order, i.e. a state
of symmetry-breaking fluctuations \cite{Fernandes2012}. Since these
fluctuations are naturally strong near the phase transition of the
primary order parameter, this line of reasoning explains the existence
of multiple nearby ordered states, largely using symmetry arguments.
It allows for predictability of complex phase diagrams, even in strongly
correlated materials.

\begin{figure*}
\begin{centering}
\includegraphics[width=\linewidth]{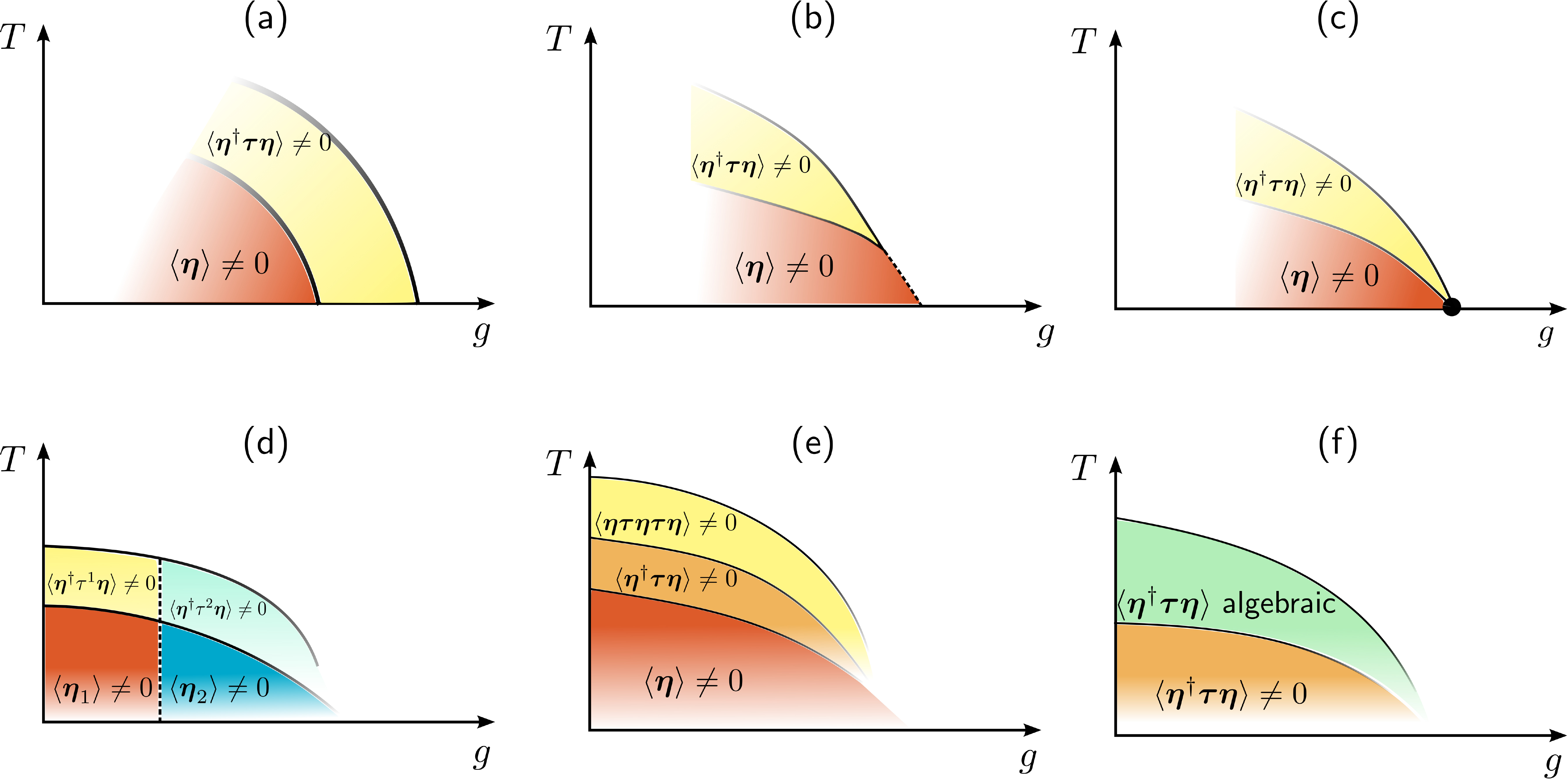}
\par\end{centering}
\caption{Schematic phases diagrams for the primary order (denoted by the parent
order parameter $\left\langle \boldsymbol{\eta}\right\rangle $) and
the vestigial order (denoted by the composite order parameters $\left\langle \boldsymbol{\eta}^{\dagger}\boldsymbol{\tau\eta}\right\rangle $
and $\left\langle \boldsymbol{\eta}\boldsymbol{\tau\eta}\boldsymbol{\tau\eta}\right\rangle $).
Second-order (first-order) transitions are denoted by solid (dashed)
lines. Panels (a), (b), and (c) show three possible outcomes for the
quantum phase transitions of the vesitigial and primary orders, in
the case when their finite-temperature phase transitions are split.
Panel (d) illustrates the appearance of two different vestigial orders
when the condensed component of the order parameter $\left\langle \boldsymbol{\eta}\right\rangle $
of the primary phase changes across the phase diagram. Panel (e) displays
a situation in which two different vestigial orders appear, corresponding
to bilinear and trilinear composites. Panel (f) illustrates the case
in which the vestigial order itself has a regime with quasi-long-range
order, giving rise to a critical vestigial phase. The parameter $g$
here corresponds to some external tuning parameter. Other phase diagrams
not shown here are also possible \label{fig_phase_diagrams}. }
\end{figure*}

The richness of the phase diagrams involving vestigial orders contrast
with the well-known phase diagrams involving competing phases \cite{Kosterlitz76,Aharony03}.
In the latter, the system displays either a bicritical or a tetracritical
point, depending on whether the competing orders phase-separate or
coexist, respectively. In contrast, several outcomes are possible
in the former case, some of which are illustrated in Fig. \ref{fig_phase_diagrams}.
A key feature is that the behavior at finite temperatures can be very
different than that at $T=0$. For instance, in the simple case of
split vestigial and primary transitions at finite temperatures, the
system may display two quantum critical points (Fig. \ref{fig_phase_diagrams}a),
a single first-order quantum phase transition (Fig. \ref{fig_phase_diagrams}b),
or even a single quantum critical point (Fig. \ref{fig_phase_diagrams}c).
Importantly, in several models more than one vestigial order appears.
Two vestigial orders can appear if the non-zero component of the primary
order parameter $\left\langle \eta_{\alpha}\right\rangle $ changes
along the phase diagram (Fig. \ref{fig_phase_diagrams}d). Moreover,
certain systems can display additional vestigial phases formed by
composite trilinear order parameters (Fig. \ref{fig_phase_diagrams}e)
or quasi-long-range ordered bilinears (Fig. \ref{fig_phase_diagrams}f).
Examples of these cases will be given throughout the review. 

Historically, fluctuation-induced composite order has played an important
role in the area of frustrated magnetism and is closely related to
the concept of order-from-disorder \cite{Villain77,Shender82,Henley88}.
The identification of an emergent, vestigial Ising degree of freedom
in a frustrated two-dimensional Heisenberg model in Ref. \cite{Chandra1990}
is a beautiful and influential example for vestigial order. More recently,
the concept of composite order played a prominent role in the explanation
of nematicity, i.e. electronically-driven rotational symmetry-breaking,
in iron-based superconductors \cite{Fang2008,Xu2009,Fernandes2010b,Fernandes2012,Fernandes2014}.
As we argue here, the applicability of this concept goes well beyond
frustrated magnetism and nematicity, opening interesting routes to
investigate unusual electronic states in unconventional superconductors
and density-wave systems. 

In order to move beyond particular examples and systems, it is important
to put the concept of composite order on formal grounds, which can
be achieved using symmetry arguments. Let the complex primary order
parameter $\eta_{\alpha}$ transform under a specific irreducible
representation $\Gamma$ of the symmetry group ${\cal G}$ of the
problem. Then the components $\alpha=1,\cdots,d_{\Gamma}$ refer to
the elements within the irreducible representation of dimension $d_{\Gamma}$.
The composite order parameter 
\begin{equation}
\phi_{m}=\sum_{\alpha\beta}\eta_{\alpha}^{*}\Lambda_{\alpha\beta}^{m}\eta_{\beta}\label{eq: composite}
\end{equation}
transforms under one of the irreducible representations $\Gamma^{m}$
that is contained in the product $\Gamma^{*}\otimes\Gamma$\cite{Hergert2018,Dresselhaus}.
Here $\Lambda_{\alpha\beta}^{m}$ is a $d_{\Gamma}\times d_{\Gamma}$-dimensional
matrix that transforms under $\Gamma^{m}$. Elementary group theoretical
arguments show that symmetry-breaking composites can only be formed
out of multi-component primary order parameters, i.e. $d_{\Gamma}>1$.
Otherwise, $\phi_{m}$ must transform under the trivial representation
and will not break a symmetry. Thus, composite order of the type Eq.\eqref{eq: composite}
requires a non-Abelian symmetry group ${\cal G}$. Fortunately, there
appears plenty of those in generic condensed-matter systems.

In the remainder of this review, we will apply and generalize such
symmetry arguments to analyze composite order that is driven by strong
fluctuations. To set the stage, we start by discussing the case of
$p$-wave unconventional superconductivity (Sec. II), followed by
the cases of density-waves on the square lattice (Sec. III) and on
the hexagonal lattice (Sec. IV). The latter have important consequences
for the phase diagrams of iron-based superconductors and graphene,
respectively. Section V discusses other examples and possible extensions
of these ideas, including an example of a system where an emergent symmetry of the ground-state leads to the absence of vestigial order.
We will demonstrate that a rich plethora of electronic states with
scalar and vector chiral order, spin-nematic order, Ising-nematic
order, time-reversal symmetry-breaking order, and critical phases
emerge out of this simple underlying principle.

\section{VESTIGIAL ORDER FROM UNCONVENTIONAL SUPERCONDUCTIVITY}
To set the stage for the next sections, we start by investigating
vestigial order in unconventional superconductors. As a specific example,
we consider a $p$-wave superconductor on a tetragonal ($d=3$) or
square ($d=2$) lattice. The amplitude $\left\langle c_{\mathbf{k}\alpha}^{\dagger}c_{-\mathbf{k}\beta}^{\dagger}\right\rangle $
of a Cooper pair that consists of one electron with momentum $\mathbf{k}$
and spin $\alpha$ and another electron with $-\mathbf{k}$ and $\beta$
is efficiently characterized in terms of the d-vector $\mathbf{d}_{\mathbf{k}}$:
\begin{equation}
\Delta_{\alpha\beta}\left(\mathbf{k}\right)=\left[\left(\mathbf{d}_{\mathbf{k}}\cdot\boldsymbol{\sigma}\right)i\sigma_{y}\right]_{\alpha\beta}.
\end{equation}
Here, $\sigma_{j}$ are Pauli matrices. The Pauli principle dictates
that the d-vector is odd under inversion, i.e. $\mathbf{d}_{-\mathbf{k}}=-\mathbf{d}_{\mathbf{k}}$,
such that the gap function is antisymmetric with respect to the exchange
of the two electrons that form the Cooper pair. In the case of a triplet
Cooper pair with $S_{z}=0$, the d-vector is parallel to the $z$-axis
and can be parametrized as:

\begin{equation}
\mathbf{d}_{\mathbf{k}}=\hat{\mathbf{z}}\left(\eta_{x}\sin k_{x}a+\eta_{y}\sin k_{y}a\right).\label{d-vector}
\end{equation}
Here, $a$ is the lattice constant in the $xy$ plane. The two complex
order parameters $\eta_{x}$ and $\eta_{y}$ thus correspond to $p_{x}$
and $p_{y}$ superconducting states, respectively. Theoretically,
$p$-wave superconductivity is expected when pairing is mediated by
the exchange of ferromagnetic fluctuations. Experimentally, material
candidates for $p$-wave superconductors include the ruthenate Sr$_{2}$RuO$_{4}$\cite{Mackenzie2003,Kallin2012}
and the doped topological insulator Cu$_{x}$Bi$_{2}$Se$_{3}$\cite{Hor2010,Kriener2011}.

\subsection{Symmetry classification}
We can build a Ginzburg-Landau expansion of the free energy $f$ in
terms of the two-component order parameter $\boldsymbol{\eta}\equiv\left(\eta_{x},\eta_{y}\right)$.
The usual form for the expansion for a system with spin-orbit coupling
and tetragonal point group $D_{4h}$ is \cite{Sigrist91} (gradient
terms are neglected for the sake of clarity): 

\begin{align}
f =\frac{r}{2}\left(\left|\eta_{x}\right|^{2}+\left|\eta_{y}\right|^{2}\right)+\frac{u}{4}\left(\left|\eta_{x}\right|^{4}+\left|\eta_{y}\right|^{4}\right) + \frac{g}{2}\left|\eta_{x}\right|^{2}\left|\eta_{y}\right|^{2}+\frac{w}{8}\left(\eta_{x}\eta_{y}^{*}+\eta_{y}\eta_{x}^{*}\right)^{2}.\label{aux_SC_free_energy-1}
\end{align}

The terms that determine the allowed ground states are the quartic
ones. For our purposes, it is therefore convenient to write $f$ in
terms of bilinears:

\begin{equation}
f=\frac{r}{2}\,\phi_{0}+\frac{\left(u+g\right)}{8}\,\phi_{0}^{2}+\frac{\left(u-g\right)}{8}\,\phi_{3}^{2}+\frac{w}{8}\,\phi_{1}^{2},\label{SC_free_energy1}
\end{equation}
where 

\begin{equation}
\phi_{m}=\sum_{\alpha\beta}\eta_{\alpha}^{*}\tau_{\alpha\beta}^{m}\eta_{\beta}\label{SC_bilinears}
\end{equation}
are the possible bilinear forms, with the Pauli matrices $\tau_{\alpha\beta}^{m}$
playing the role of the matrices $\Lambda_{\alpha\beta}^{m}$ in Eq.\eqref{eq: composite}.
The absence of the $\phi_{2}^{2}$ term is a consequence of the Fierz
identity, $\phi_{1}^{2}+\phi_{3}^{2}=\phi_{0}^{2}-\phi_{2}^{2}$,
which implies that one can always express one of the allowed bilinear
forms in terms of the others. The different possible $p$-wave superconducting
states are obtained by minimizing the free energy, and are given by
\begin{eqnarray}
\boldsymbol{\eta}_{B_{1g}} & \propto & \left(1,0\right)\:{\rm or}\:\left(0,1\right),\nonumber \\
\boldsymbol{\eta}_{B_{2g}} & \propto & \left(1,\pm1\right),\nonumber \\
\boldsymbol{\eta}_{A_{2g}} & \propto & \left(1,\pm i\right).
\end{eqnarray}

Which state is realized depends on the values of the quartic coefficients.
The $B_{1g}$ superconducting state is the ground state when $u-g<\mathrm{min}\left(0,w\,\right)$;
the $B_{2g}$ ground state takes place when $u-g>w$ and $w<0$; and
the $A_{2g}$ ground state is realized when $u-g>0$ and $w>0$. 

\begin{figure}
\begin{centering}
\includegraphics[width=\linewidth]{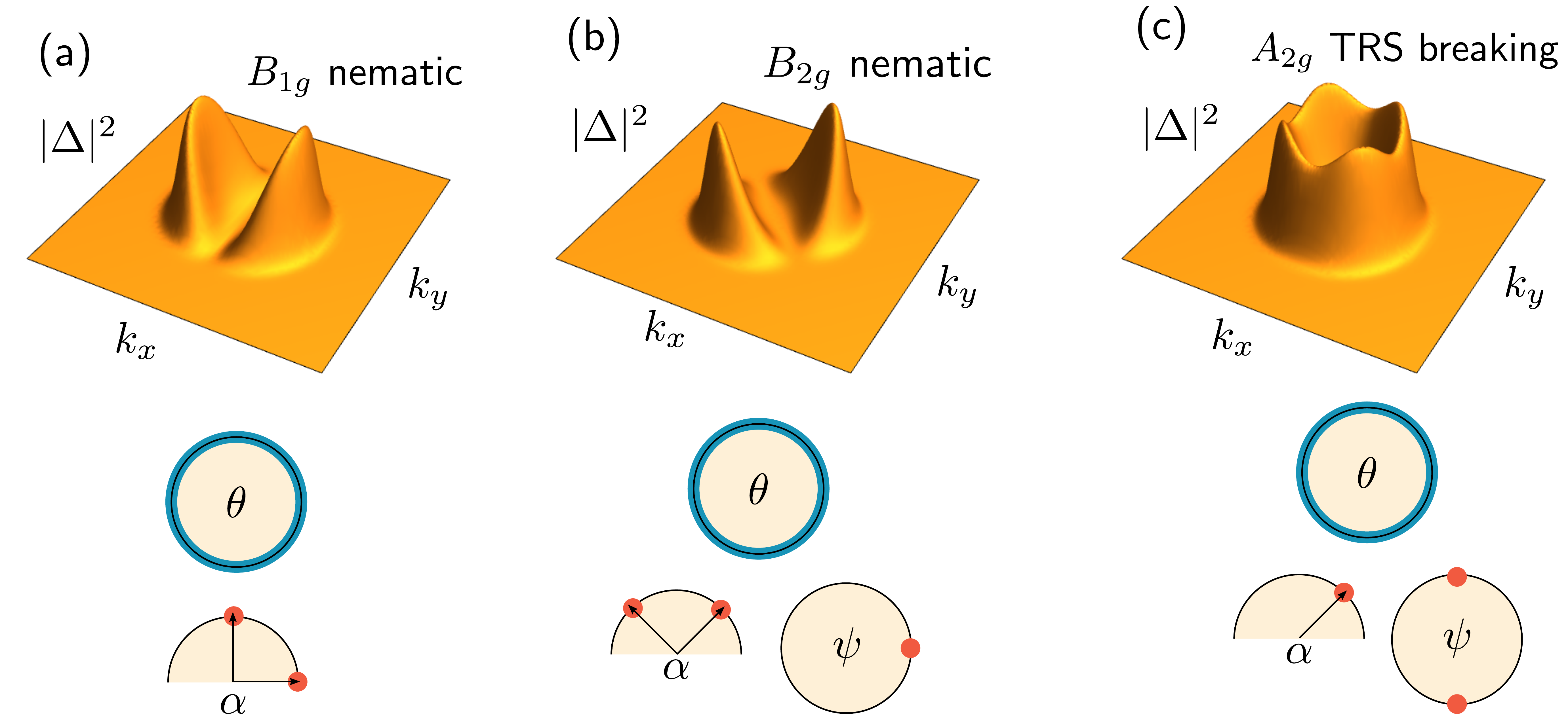}
\par\end{centering}
\caption{The three possible $S_{z}=0$ triplet superconducting states in a
tetragonal system (upper panels), and the corresponding vestigial
phases (lower panels). Panel (a) shows the $B_{1g}$-nematic state;
panel (b), the $B_{2g}$-nematic state; and panel (c), the $A_{2g}$
time-reversal symmetry-breaking state. The primary phases are illustrated
by three-dimensional plots of the gap function $\left|\Delta\right|^{2}$,
defined in Eq. (\ref{d-vector}), around a circular Fermi surface.
The vestigial phases are illustrated by the behavior of the phases
defined in Eq. (\ref{gap_eta}). While the global phase $\theta$
is always fluctuating in the vestigial phases, one of the phases $\alpha$
or $\psi$ can acquire two different values, evidencing the Ising-like
character of the vestigial order parameters. \label{fig_p_wave}}

\end{figure}

At the superconducting transition, the global $U\left(1\right)$ symmetry
is broken or, in the case of a two-dimensional system, algebraic order
sets in via a Berezinskii-Kosterlitz-Thouless (BKT) transition. In
addition, each of the three possible ground states also breaks a discrete
Ising-like ($Z_{2}$) symmetry of the system, as illustrated schematically
in Fig. \ref{fig_p_wave}. The $B_{1g}$ state breaks the tetragonal
symmetry of the lattice such that the $x$ and $y$ directions are
inequivalent. This is shown in the upper panel of Fig. \ref{fig_p_wave}a,
which plots $\left|\Delta\right|^{2}$ for a circular Fermi surface.
It therefore corresponds to a $B_{1g}$ nematic superconductor, since
the system retains horizontal and vertical mirror symmetries. The
$B_{2g}$ state with $\eta_{x}=\pm\eta_{y}$ also breaks the tetragonal
symmetry of the lattice, but by making the diagonal directions $xy$
and $x\bar{y}$ inequivalent (upper panel of Fig. \ref{fig_p_wave}b).
As a result, it is a $B_{2g}$ nematic superconductor, where the diagonal
mirror symmetries are preserved. Finally, the state with $\eta_{x}=\pm i\eta_{y}$
breaks time-reversal symmetry (upper panel of Fig. \ref{fig_p_wave}c).
It supports orbital currents associated with the chirality of the
gap function, and as such it transforms as the $A_{2g}$ irreducible
representation of the tetragonal group.

These properties of a $p$-wave superconductor are efficiently captured
within the framework outlined in the introduction. The symmetry group
of the problem is $\mathcal{G}=D_{4h}\otimes U(1)$, where $D_{4h}$
refers to the tetragonal point group that describes the square and
tetragonal lattices, and $U(1)$ is the continuous group related to
the complex nature of the order parameters $\eta_{\alpha}$. Importantly,
the $\eta_{x}$ and $\eta_{y}$ order parameters transform according
to the two-dimensional irreducible representation $E_{u}$ of $D_{4h}$,
i.e. $\Gamma=E_{u}\times e^{im\theta}$ . We focus on composite order
parameters that do not break the $U\left(1\right)$ order parameter.
The product 
\begin{equation}
\Gamma^{*}\otimes\Gamma =  E_{u}\otimes E_{u}  = A_{1g}\oplus B_{1g}\oplus B_{2g}\oplus A_{2g}\label{eq:Eu_irreps}
\end{equation}
is then decomposed in terms of $4$ one-dimensional irreducible representations
of the $D_{4h}$ group. From Eq. (\ref{eq:Eu_irreps}), we conclude
that there are four different values of the irreducible representation
index, $m=0,\,1,\,2,\,3$. The associated bilinear forms are the same
as those introduced in Eq. \eqref{SC_bilinears}. In explicit form,
we have: $\phi_{0}  =\boldsymbol{\eta}^{\dagger}\tau^{0}\boldsymbol{\eta}=\left|\eta_{x}\right|^{2}+\left|\eta_{y}\right|^{2}$, $\phi_{1} =\boldsymbol{\eta}^{\dagger}\tau^{1}\boldsymbol{\eta}=\eta_{x}\eta_{y}^{*}+\eta_{y}\eta_{x}^{*}$, $\phi_{2}  =\boldsymbol{\eta}^{\dagger}\tau^{2}\boldsymbol{\eta}=i\left(\eta_{x}\eta_{y}^{*}-\eta_{y}\eta_{x}^{*}\right)$, and $\phi_{3}  =\boldsymbol{\eta}^{\dagger}\tau^{3}\boldsymbol{\eta}=\left|\eta_{x}\right|^{2}-\left|\eta_{y}\right|^{2}$. 
Because $\phi_{0}$ transforms as the trivial irreducible representation
$A_{1g}$, it does not break any symmetry of the system. As a result,
it cannot serve as a vestigial order parameter, but instead corresponds
to fluctuations present in the vicinities of the normal-state to superconducting
phase transition, regardless of the nature of the superconducting
state. $\phi_{1}$, on the other hand, transforms as the $B_{2g}$
irreducible representation and, as such, is a nematic vestigial order
parameter that breaks the tetragonal symmetry of the system. It is
clear that it is only compatible with the $B_{2g}$ nematic superconducting
ground state, in which $\eta_{x}=\pm\eta_{y}$. Similarly, $\phi_{3}$
transforms as the $B_{1g}$ irreducible representation, and is thus
also a nematic vestigial order, compatible with the $B_{1g}$ nematic
superconducting state in which either $\eta_{x}=0$ or $\eta_{y}=0$.
Finally, $\phi_{2}$ transforms as the $A_{2g}$ irreducible representation,
and thus breaks time-reversal symmetry, since $A_{2g}$ corresponds
to orbital angular momentum along the $z$ axis. Clearly, it is only
compatible with the time-reversal symmetry-breaking superconducting
ground state, in which $\eta_{x}=\pm i\eta_{y}$. 

The coefficients of the terms that are quadratic in $\phi_{i}^{2}$
(with $i=1,\,2,\,3$), also called ``masses'' in field theory, determine
which of the vestigial orders can appear. From Eqs. (\ref{SC_free_energy1})
and the Fierz identity we conclude that if $u-g<\mathrm{min}\left(0,w\,\right)$,
the mass of the $\phi_{3}^{2}$ term is negative, and smaller than
the masses of the $\phi_{1}^{2}$ and $\phi_{2}^{2}$ terms, indicating
a tendency towards $B_{1g}$ vestigial order. $u-g<\mathrm{min}\left(0,w\,\right)$
is also the condition that ensures that the ground state is the $B_{1g}$
nematic superconducting state. Similar results hold in the other two
regions of the parameter space $(u-g,\,w)$. 

The key remaining question is whether the composite order parameter
$\phi_{i}$ can condense even in the absence of superconducting order,
i.e. whether the system can display a regime in which $\left\langle \boldsymbol{\eta}^{\dagger}\tau^{i}\boldsymbol{\eta}\right\rangle \neq0$
but $\left\langle \boldsymbol{\eta}\right\rangle =0$. The $\phi_{i}$
are $Z_{2}$ (Ising-like) order parameters in this case, since they
each transform according to one-dimensional irreducible representations,
whereas $\boldsymbol{\eta}$ are complex $U(1)$ fields. Within mean-field,
both the $Z_{2}$ and $U(1)$ symmetries are broken at the same temperature.
However, once fluctuations are included, the natural result is that
they are broken at two different temperatures or that a joint first
order transition takes place. These two options are the generic behaviors
of two order parameters that break different symmetries, whereas the
simultaneous and second-order transition is only correct within a
mean-field description. Since mean-field theory is appropriate for
many superconductors, one expects quantitatively small effects. There are, however, a number of low-density and low-dimensional
superconductors that are governed by sizable fluctuations of the superconducting
order parameter and that are strong candidates for vestigial order. Examples are doped Bi$_{2}$Se$_{3}$\cite{Hor2010,Kriener2011}
and the half-Heusler systems LuPtBi and YbPtBi\cite{Goll2008,Butch2011,Tafti2013,Xu2014}.
In fact, the observed nematic order below $T_{c}$ in Cu- and Sr\textendash doped
Bi$_{2}$Se$_{3}$ \cite{Matano2016,Yonezawa2017,Pan2016,Du2017,Asaba2017,Shen2017}
strongly suggests a nematic phase above $T_{c}$\cite{Hecker2018}.
Due to the trigonal point group $D_{3d}$ of this material, it follows
that the vestigial nematic order parameter behaves like a three-state
Potts model\cite{Hecker2018}. The cuprates are another class of materials
where strong superconducting fluctuations are present. However, the
gap function is $d_{x^{2}-y^{2}}$, which transforms as a one-dimensional
irreducible representation of the $D_{4h}$ group. Consequently, vestigial
order related to superconductivity in the cuprates can only arise
if there is additional translational symmetry breaking, as is the
case for pair-density-wave states. Several recent works have focused
on the issue of vestigial orders of the pair-density-waves, mostly
in the context of the cuprates \cite{Agterberg08,Berg09,Yuxuan15,Zaanen17}.

\subsection{Model calculations}
\label{sec:model-calculations}
Symmetry arguments can take us this far, but to proceed and determine
whether the superconducting and vestigial orders are split, explicit
calculations are necessary. Approaching the vestigial order from the
melted ordered state, we parametrize the order parameter in terms
of:
\begin{equation}
\boldsymbol{\eta}\left(\mathbf{x}\right)=\sqrt{n_{0}}e^{i\theta\left(\mathbf{x}\right)}\left(\begin{array}{c}
\cos\alpha\left(\mathbf{x}\right)\\
e^{i\psi\left(\mathbf{x}\right)}\sin\alpha\left(\mathbf{x}\right)
\end{array}\right),\label{gap_eta}
\end{equation}
with constant $n_{0}$. There are three coordinate-dependent phase
variables: the global phase $\theta$, the relative phase $\psi$
between the two $p$-wave components $p_{x}$ and $p_{y}$, and the
phase $\alpha$ that selects whether both components are simultaneously
present. In each of the three vestigial phases, $\left\langle e^{i\theta\left(\mathbf{x}\right)}\right\rangle =0$,
implying that the system has no superconducting order. Moreover, either
$\alpha$ or $\psi$ can acquire two values in a given vestigial phase,
highlighting the Ising character of the composite order parameters
(see Fig. \ref{fig_p_wave})

For concreteness, let us consider a two-dimensional system with $u-g>w>0$,
which corresponds to the mean-field ground state $\boldsymbol{\eta}_{A_{2g}}\propto\left(1,\pm i\right)$.
The effective action $S\equiv F/T$, where $F$ is the total free
energy has two contributions: the gradient term (we neglect here
the coupling to the electromagnetic field)
\begin{equation}
S_{{\rm grad}} =  \frac{1}{2T}\int d^{2}x\left\{ \left(\partial_{\mu}\theta\right)^{2}+\left(\partial_{\mu}\alpha\right)^{2}+\sin^{2}\alpha\left(\partial_{\mu}\psi\right)^{2}\right.  +  \left.2\sin^{2}\alpha\partial_{\mu}\psi\partial_{\mu}\theta\right\} 
\end{equation}
with some dimensionless temperature $T$, and the potential term
\begin{equation}
S_{{\rm pot}}=-\frac{\Delta}{a^{2}T}\int d^{2}x\sin^{2}\left(2\alpha\right)\sin^{2}\psi,
\end{equation}
where $\Delta=wn_{0}^{2}a^{2}$ is a dimensionless constant with $w>0$
from Eq.\eqref{aux_SC_free_energy-1}. This action can be analyzed
using renormalization-group techniques (for a related problem, see
Ref. \cite{Fellows2012}). The key result is the onset of time-reversal symmetry (TRS) breaking at a temperature $T_{0}\sim2\pi/\log\left(\Delta^{-1}\right)$. For $T>T_{0}$,
the gradient term dominates the renormalization-group flow, and the
system behaves similarly to Heisenberg $O(3)$ spins. In this regime,
the superconducting correlation length follows the usual behavior
of the non-linear sigma model with spin correlation length $\xi\left(T>T_{0}\right)=ae^{\pi/T}$. 

Below $T_{0}$, the potential term starts to dominate. Because $\Delta>0$
increases under the renormalization group flow, the effect of this
term is to lock the variables $\alpha$ and $\psi$ in order to minimize
the energy, i.e. $\alpha=\frac{\pi}{4}$ and $\psi=\frac{\pi}{2}$
or $\psi=\frac{3\pi}{2}$ . As a result, the order parameter is that
of a $p_{x}\pm ip_{y}$ superconductor with a fluctuating phase:
\begin{equation}
\boldsymbol{\eta}\left(x\right)=\sqrt{\frac{n_{0}}{2}}e^{i\theta\left(x\right)}\left(\begin{array}{c}
1\\
\pm i
\end{array}\right),
\end{equation}
Now the only relevant variable is the overall superconducting phase,
such that the gradient term becomes 
\begin{equation}
S_{{\rm grad}}\rightarrow\frac{1}{2T}\int d^{2}x\left(\partial_{\mu}\theta\right)^{2}
\end{equation}
which is the same action as the usual XY-model. As a result, the system
becomes governed by the Berezinskii-Kosterlitz-Thouless (BKT) behavior
of the XY-model with the key difference that the size of the vortex
core is $\xi\left(T_{0}\right)\approx\frac{a}{\sqrt{\Delta}}$. 

Because $\xi\left(T_{\mathrm{BKT}}\right)\rightarrow\infty$, the
BKT transition temperature $T_{\mathrm{BKT}}$ is clearly below $T_{0}$,
even though we find, following Ref.\cite{Fellows2012}, that both
temperatures are parametrically of the same order. To unveil the meaning
of the temperature $T_{0}$, we note that the potential term can be
alternatively expressed in terms of the vestigial $A_{2g}$ order
parameter $\phi_{2}$
\[
S_{{\rm pot}}=-\frac{\Delta}{a^{2}T}\int d^{2}x\left(\frac{\phi_{2}}{n_{0}}\right)^{2}
\]

As the correlation length increases, regions of typical size $\xi$
essentially share the same value of the Ising variable $\phi_{2}/n_{0}\approx\pm1$.
For a two-dimensional system, this implies that a true Ising-like
phase transition takes place when the correlation length becomes comparable
to the Ising domain-wall thickness $a/\sqrt{\Delta}$:
\begin{equation}
\Delta\frac{\xi\left(T\right)^{2}}{a^{2}}\approx1.\label{eq:T0 d2}
\end{equation}

This immediately yields $T_{{\rm c,Ising}}=T_{0}$. Thus, $T_{0}>T_{\mathrm{BKT}}$
signals a true Ising-like phase transition to the vestigial state
that breaks time-reversal symmetry, but does not display quasi-long-range
superconducting order. This result agrees with analyses of related
models that also find an Ising order onsetting above the BKT transition
\cite{Korshunov02,Vicari05}. 

One can also approach the vestigial phase if the system is not exactly
two-dimensional, i.e. if true superconducting long-range order can
take place. Generally, different techniques can be employed, such
as the renormalization-group \cite{Qi09,Millis10,Fernandes2012},
self-consistent Gaussian approximation \cite{Nie17}, and the saddle-point
large-$N$ approximation \cite{Fernandes2012}. For the specific case
of a $p$-wave superconductor, the self-consistent Gaussian approximation
was employed in Ref. \cite{Fischer16}. Here, we will focus on the
large-$N$ approach: in this method, one starts with the free energy
(\ref{SC_free_energy1}), complemented by the standard gradient terms,
and decouples the quartic coefficients (quadratic in the bilinears)
using Hubbard-Stratonovich transformations. In the parameter regime
relevant for $A_{2g}$ superconducting order, $u-g>w>0$, it is sufficient
to keep only the fields corresponding to the $\phi_{0}$ and $\phi_{2}$
bilinears, obtaining the action

\begin{align}
S =\int_{k}\boldsymbol{\eta}_{k}^{\dagger}\left[\left(r+\phi_{0}+k^{2}\right)\boldsymbol{\tau}_{0}+\phi_{2}\boldsymbol{\tau}_{y}\right]\boldsymbol{\eta}_{k}  +\frac{\phi_{2}^{2}}{4w}-\frac{\phi_{0}^{2}}{4\left(u+g+w\right)}.
\end{align}

In the disordered state, the fields $\boldsymbol{\eta}$ are fluctuating
and can be integrated out exactly, resulting in an action that depends
only on $\phi_{0}$ and $\phi_{2}$. The equation of state for $\phi_{2}$
can then be obtained using a saddle-point approximation, which is
formally exact in the limit where the number $N$ of components of
$\boldsymbol{\eta}$ is $N\rightarrow\infty$. To linear order in
the vestigial order parameter $\phi_{2}$, one obtains:

\begin{equation}
\frac{\phi_{2}}{w}=A\xi^{4-d}\phi_{2}
\end{equation}
where $A$ is some constant, $\xi\left(T\right)$ is the temperature-dependent
correlation length, and $d>2$ is the dimensionality of the system.
This equation allows a non-zero $\phi_{2}$ value at the critical
temperature $T_{0}$ where $\xi\left(T_{0}\right)=\left(Aw\right)^{-\frac{1}{4-d}}$,
which takes place before the temperature $T_{c}$ in which long-range
superconductivity appears, since $\xi\left(T_{c}\right)\rightarrow\infty$.
For $d=2$ we recover the previous result for the correlation
length at the vestigial transition. The only way to avoid vestigial
order at a separate transition temperature $T_{0}>T_{c}$ is via a
simultaneous first-order transition, since in this case $\xi(T_{c})$
no longer diverges. This is the generic behavior that occurs in isotropic,
three-dimensional systems\cite{Fernandes2012}. The split second-order
transition in low-dimensional and anisotropic three-dimensional systems
is a consequence of the enhanced role of fluctuations of the primary
order parameter\cite{Karahasanovic2016}. The quantum dynamics near
$T=0$, however, may place the system closer to its upper critical
dimension, thus reducing the impact of fluctuations and favoring a
single first-order quantum transition \cite{Fernandes13} (see Figs.
\ref{fig_phase_diagrams}a and b).

These analyses reveal that there can be no single second-order phase
transition into a multi-component superconductor: either there are
two separate transitions or a single first-order transition. This
simple yet robust result has important implications for the interpretation
of experimental data on material candidates for $p$-wave superconductivity
(see also Ref. \cite{Fischer16}). The same behavior holds also for
any superconducting state with a multi-component order parameter that
transforms according to any of the 32 point groups of three-dimensional
crystalline systems. Specifically, vestigial orders originating from
superconductivity are possible for the 15 point groups with higher-dimensional
irreducible representations, i.e. for all the cubic groups $T$, $T_{h}$,
$T_{d}$, $O$, and $O_{h}$, the tetragonal groups $C_{4v}$, $D_{2d}$,
$D_{4}$, and $D_{4h}$, the hexagonal groups $C_{6v}$, $D_{3h}$,
and $D_{6h}$, and the trigonal groups $C_{3v}$, $D_{3d}$, and $D_{3}$.
On the other hand, no vestigial order of translationally-invariant
superconducting states occurs in an orthorhombic, monoclinic, or triclinic
system.

\section{VESTIGIAL ORDER FROM DENSITY-WAVES IN THE SQUARE LATTICE}
\label{sec:Vestigial-order-from}

\begin{figure}
\begin{centering}
\includegraphics[width=\linewidth]{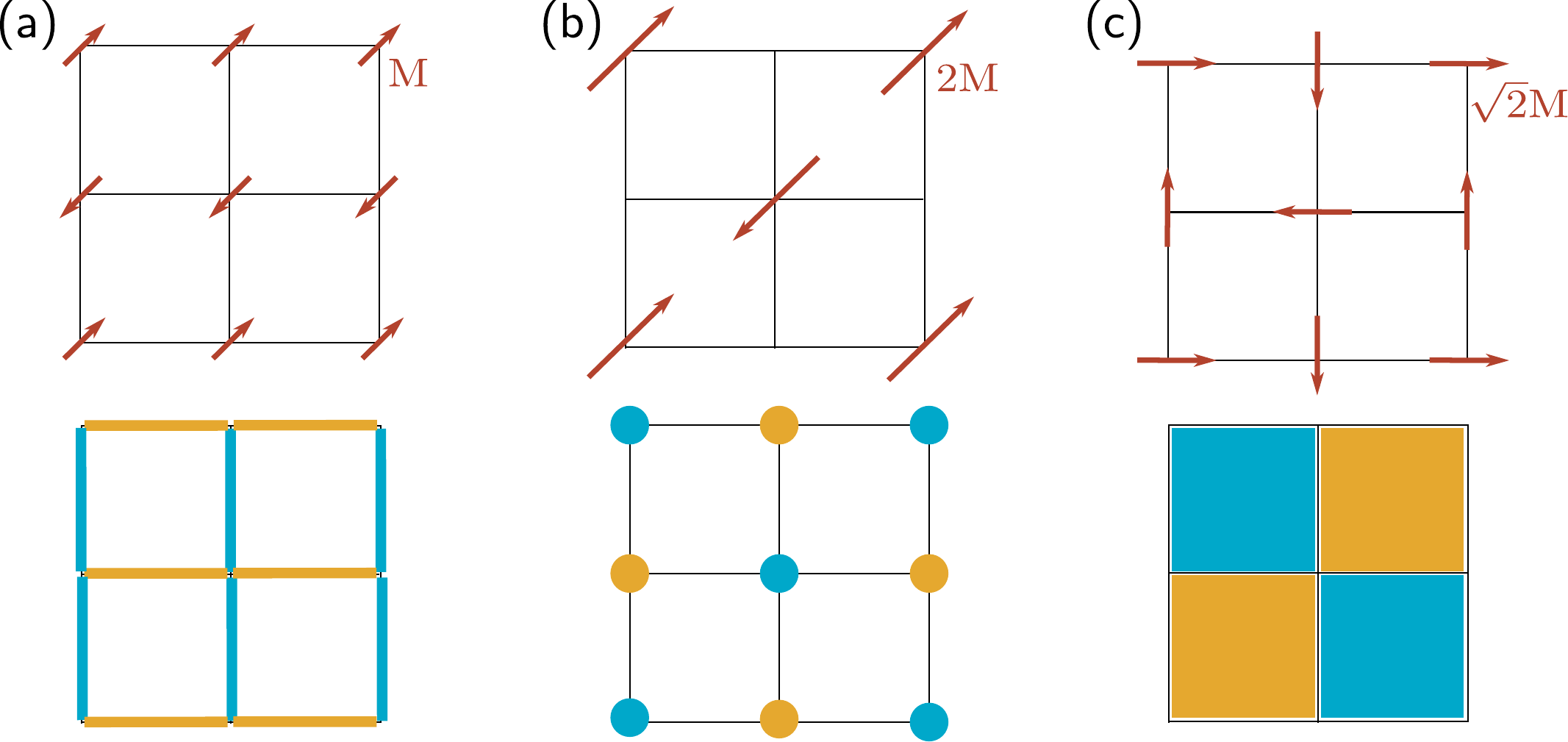}
\par\end{centering}
\caption{Schematic representation of the three possible square-lattice spin
density-wave ground states (upper panels) with ordering vectors $\mathbf{Q}_{1}=\left(\pi,0\right)$
and $\mathbf{Q}_{2}=\left(0,\pi\right)$, and their corresponding
vestigial phases (lower panels). Panel (a) refers to the single-\textbf{Q
}$C_{2}$-symmetric magnetic phase and its corresponding nematic vestigial
phase, characterized by unequal bonds. Panel (b) shows the collinear
double-\textbf{Q }$C_{4}$-symmetric magnetic phase and its corresponding
charge-ordered vestigial phase, characterized by unequal sites. Note that, in the magnetically ordered state, half of the sites have zero magnetization. Panel
(c) illustrates the non-collinear double-\textbf{Q }$C_{4}$-symmetric
magnetic phase and its corresponding spin-current vestigial phase,
characterized by unequal plaquettes. As indicated in the figure, the magnitudes of the local magnetization are different in each ordered state. \label{fig_square_lattice}}

\end{figure}

We now proceed to apply the formalism developed above to classify
possible vestigial orders arising from density-waves on the square
lattice. We start with the richer case of spin density-waves. As explained,
the ground state must be degenerate in order for non-trivial composite
operators to emerge. The standard N\'eel-like order, with wave-vector
$\mathbf{Q}=\left(\pi,\pi\right)$, does not support vestigial orders
that break the point group symmetry of the lattice. The simplest non-trivial
case is then that of two degenerate magnetic ground states that are
related by a symmetry of the lattice, corresponding to two ordering
vectors $\mathbf{Q}_{1}=\left(\pi,0\right)$ and $\mathbf{Q}_{2}=\left(0,\pi\right)$.
The local spin can then be written as
\begin{equation}
\mathbf{S}\left(\mathbf{r}\right)=\mathbf{m}_{1}\cos{\left(\mathbf{Q}_{1}\cdot\mathbf{r}\right)}+\mathbf{m}_{2}\cos{\left( \mathbf{Q}_{2}\cdot\mathbf{r} \right)},
\end{equation}
where $\mathbf{m}_{a}$ are the real vector order parameters associated
with $\mathbf{Q}_{a}$, where $a=1,\,2$. In the square lattice, there
are three possible magnetic ground states \cite{Lorenzana08,Fernandes2016},
illustrated in Fig. \ref{fig_square_lattice}: a $C_{2}$-symmetric
single-\textbf{Q }spin density-wave, corresponding to only one $\mathbf{m}_{a}$
being non-zero; a $C_{4}$-symmetric collinear double-\textbf{Q} spin
density-wave, corresponding to $\mathbf{m}_{1}\parallel\mathbf{m}_{2}\neq0$;
and a $C_{4}$-symmetric non-collinear double-\textbf{Q }spin density-wave,
corresponding to $\mathbf{m}_{1}\perp\mathbf{m}_{2}\neq0$.

This description has been widely employed to discuss nematicity and
magnetism in iron-based materials \cite{Lorenzana08,Eremin10,Brydon11,Giovannetti11,Fernandes2012,Wang15,Fernandes2016}.
It arises from either a $J_{1}$-$J_{2}$ localized spin model or
an itinerant microscopic model with partially nested Fermi pockets
\cite{Fang2008,Xu2009,Lorenzana08,Fernandes2012}. In what follows,
we will not repeat arguments that were extensively presented elsewhere
\cite{Fernandes2014,Fernandes2012b}, but instead give a symmetry-based
analysis of the allowed vestigial states.

We start by writing down the symmetry group of the problem without
spin-orbit interaction:
\begin{equation}
{\cal G}=C_{4v}^{'''}\times SO\left(3\right).
\end{equation}
Here, $C_{4v}^{'''}$ is called the extended point group \cite{Serbyn13,Venderbos16}.
It corresponds to the standard point group $C_{4v}$ supplemented
by three translations: $T_{1}=\left(1,0\right)$, $T_{2}=\left(0,1\right)$,
and $T_{3}=\left(1,1\right)$. It is convenient to consider this group
because the density-wave order parameters break translational symmetry.
We do not include inversion symmetry explicitly here. Because we have
a vector order parameter, it transforms under the irreducible representation
$\Gamma=E_{5g}\otimes\Gamma^{S=1}$. Since $d_{E_{5g}}=2$ and $d_{S}=2S+1$,
the dimensionality of the irreducible representation of the primary
order parameter is $d_{\Gamma}=2\times\left(2+1\right)=6$. As a result,
the order parameter can be written as $\eta_{A}=\left(\mathbf{m}_{1},\mathbf{m}_{2}\right),$where
the $\mathbf{m}_{a}$ are three-component vectors in spin space. Note
that such a classification of the primary order parameters was done
in Ref. \cite{Venderbos16}, from which we borrow the group-theory
notation. Here, our goal is to systematically discuss the possible
vestigial orders.

The bilinear forms can be analyzed by using the following results
\begin{equation}
E_{5}\otimes E_{5}=A_{1}\oplus B_{2}^{'}\oplus A_{2}^{'}\oplus B_{1}
\end{equation}
and 
\begin{equation}
\Gamma^{S*}\otimes\Gamma^{S}=\bigoplus_{j=0}^{2S}\Gamma^{j}\label{eq_Gamma_s}
\end{equation}

The primes in $A'_{2}$ and $B'_{2}$ indicate that translational
symmetry is broken by $T_{3}=(1,1)$. We thus obtain:

\begin{equation}
\Gamma^{*}\otimes\Gamma=\left(A_{1}\oplus B_{2}^{'}\oplus A_{2}^{'}\oplus B_{1}\right)\otimes\left(\Gamma^{0}\oplus\Gamma^{1}\oplus\Gamma^{2}\right)
\end{equation}

The index of a irreducible representation of the product, $m=\left(r,j\right)$,
is then a combination of the four spatial irreducible representations
$r=\left(0,1,2,3\right)=\left(A_{1},B_{2}^{'},A_{2}^{'},B_{1}\right)$
and the spin $j$. As a result, the possible composite operators are
written as:

\begin{equation}
\phi_{m\equiv\left(r,j\right)}^{\mu}=\sum_{A,B}\eta_{A}\Lambda_{A,B}^{m,\mu}\eta_{B}
\end{equation}
with matrices 
\begin{equation}
\Lambda_{A,B}^{m\equiv(r,j),\mu}=\tau_{ab}^{r}\lambda_{\alpha\beta}^{j,\mu}.
\end{equation}

These matrices transform according to one of the irreducible representations
$\Gamma^{m}$ contained in the product $\Gamma^{*}\otimes\Gamma$.
The number of matrices is given by the dimensionality $d_{m}$ of
$\Gamma^{m}$, so that the index $\mu=1,\ldots,d_{m}$. The indices
$A=\left(a,\alpha\right)$, $B=(b,\beta)$ combine point and spin
group indices, such that $\eta_{A}\equiv m_{a}^{\alpha}$. The $\tau_{ab}^{r}$
are the unit matrix $\tau^{0}$ and the three Pauli matrices $\tau^{r}$.
The $3\times3$ matrices $\lambda^{j}$ for $j=0\text{, }\text{1},$
and $2$ act in spin space and are given as follows: For $j=0$ we
have $\lambda_{\alpha\beta}^{0,0}=\delta_{\alpha\beta}$, and the
composite order parameter can be expressed as a scalar $\phi_{(r,j=0)}$.
For $j=1$ we have three matrices $\lambda_{\alpha\beta}^{1,\mu}=i\epsilon_{\alpha\beta\mu}$
corresponding to the three anti-symmetric Gell-Mann matrices\textbf{\emph{.
}}Thus, we can express the composite order parameter as a vector $\boldsymbol{\phi}_{(r,j=1)}$.
Finally for $j=2$, we use the five symmetric Gell-Mann matrices.
They can be labelled by a double index $\left(\mu,\mu'\right)$ of
a symmetric tensor, where $\mu$ and $\mu'$ take three values each:
\begin{equation}
\lambda_{\alpha\beta}^{2,\left(\mu,\mu'\right)}=\frac{1}{2}\left(\delta_{\alpha\mu}\delta_{\beta\mu'}+\delta_{\alpha\mu'}\delta_{\beta\mu}\right)-\frac{1}{3}\delta_{\alpha\beta}\delta_{\mu\mu'}.
\end{equation}
In this case, the vestigial order parameter is a second-rank tensor
$\phi_{(r,j=2)}^{\mu\mu'}$. This exhausts all $3\times3$ matrices,
which is what we expect for an order parameter that transforms as
$S=1.$

We first consider $j=0$. There are three possible non-vanishing scalar
bilinears
\begin{eqnarray}
\phi_{\left(0,0\right)} & = & \mathbf{m}_{1}\cdot\mathbf{m}_{1}+\mathbf{m}_{2}\cdot\mathbf{m}_{2}\nonumber \\
\phi_{\left(1,0\right)} & = & 2\mathbf{m}_{1}\cdot\mathbf{m}_{2}\nonumber \\
\phi_{\left(3,0\right)} & = & \mathbf{m}_{1}\cdot\mathbf{m}_{1}-\mathbf{m}_{2}\cdot\mathbf{m}_{2},
\end{eqnarray}

Note that $\phi_{\left(2,0\right)}$ vanishes, since the $\mathbf{m}_{i}$
are real vectors. While $\phi_{\left(0,0\right)}$ transforms trivially
($A_{1}$ representation), we obtain two vestigial order parameters
that break spatial symmetries, without breaking spin-space symmetries.
$\phi_{\left(3,0\right)}$, which transforms as $B_{1}$, is an Ising-nematic
order parameter, which is frequently observed in iron-based systems
(see Fig. \ref{fig_square_lattice}a). It is the vestigial phase of
the single-\textbf{Q }magnetic ground state. $\phi_{\left(1,0\right)}$,
which transforms as $B'_{2}$, corresponds to a scalar that breaks
translational symmetry (with ordering vector $\mathbf{Q}_{1}+\mathbf{Q}_{2}=\left(\pi,\pi\right)$),
while preserving the tetragonal symmetry of the lattice (see Fig.
\ref{fig_square_lattice}b). It thus corresponds to a checkerboard
charge order, and is the vestigial phase of the $C_{4}$-symmetric
collinear double-\textbf{Q} magnetic state observed in several iron-based
systems\cite{Kim2010,Hassinger2012,Avci2014,Wang2016,Boehmer2015,Allred2015,Hassinger2016,Allred2016}.
Interestingly, in the phase diagrams of these compounds, the single-\textbf{Q
}phase undergoes a transition to the double-\textbf{Q }phase as function
of doping. A little explored problem is the interplay between the
two corresponding vestigial orders in this case where the primary
order itself changes (see Fig. \ref{fig_phase_diagrams}d). 

For $j=1$, the only non-zero vestigial order is the vector composite
order parameter
\begin{equation}
\mathbf{\boldsymbol{\phi}}_{\left(2,1\right)}=2\mathbf{m}_{1}\times\mathbf{m}_{2}.
\end{equation}

It corresponds to a vector chirality, which is manifested as spin
current loops that are staggered between different plaquettes, forming
an imaginary spin density-wave with ordering vector $\mathbf{Q}_{1}+\mathbf{Q}_{2}=\left(\pi,\pi\right)$
(see Fig. \ref{fig_square_lattice}c). It is the vestigial phase of
the $C_{4}$-symmetric non-collinear double-\textbf{Q} magnetic state
observed recently in doped CaKFe$_{4}$As$_{4}$\cite{Meier2018}.
The three non-trivial states $\phi_{\left(1,0\right)}$, $\phi_{\left(3,0\right)}$,
and $\mathbf{\boldsymbol{\phi}}_{\left(2,1\right)}$ were recently
discussed in Ref. \cite{Fernandes2016} and analyzed using a large-$N$
approximation. In a two-dimensional system, where long-range order
of the primary order parameters is prohibited by the Hohenberg-Mermin-Wagner
theorem, a vestigial phase having only the Ising-like order parameters
$\phi_{\left(1,0\right)}$ or $\phi_{\left(3,0\right)}$ will take
place. Note that the continuous composite order parameter $\mathbf{\boldsymbol{\phi}}_{\left(2,1\right)}$
cannot condense in a two-dimensional system; in Ref. \cite{Fernandes2016},
it was argued that a vestigial phase with $\mathbf{\boldsymbol{\phi}}_{\left(2,1\right)}\neq0$
but $\mathbf{m}_{i}=0$ is possible in strongly anisotropic three-dimensional
systems. 

Finally, there is also the possibility for three more vestigial states
with $j=2$: 

\begin{align}
\phi_{\left(0,2\right)}^{\mu\mu'} & =  m_{1}^{\mu}m_{1}^{\mu'}+m_{2}^{\mu}m_{2}^{\mu'} -  \frac{1}{3}\delta_{\alpha\beta}\left(\mathbf{m}_{1}\cdot\mathbf{m}_{1}+\mathbf{m}_{2}\cdot\mathbf{m}_{2}\right)\nonumber \\
\phi_{\left(1,2\right)}^{\mu\mu'} & =  m_{1}^{\mu}m_{2}^{\mu'}+m_{2}^{\mu}m_{1}^{\mu'} -  \frac{1}{3}\delta_{\mu\mu'}\left(\mathbf{m}_{1}\cdot\mathbf{m}_{2}+\mathbf{m}_{2}\cdot\mathbf{m}_{2}\right)\nonumber \\
\phi_{\left(3,2\right)}^{\mu\mu'} & =  m_{1}^{\mu}m_{1}^{\mu'}-m_{2}^{\mu}m_{2}^{\mu'}  -  \frac{1}{3}\delta_{\mu\mu'}\left(\mathbf{m}_{1}\cdot\mathbf{m}_{1}-\mathbf{m}_{2}\cdot\mathbf{m}_{2}\right).\label{eq_tensor_rank2}
\end{align}

While $\phi_{\left(0,2\right)}^{\mu\mu'}$ (with $\mu,\mu'=1,2,3$)
corresponds to pure spin-nematicity (i.e. nematic order in spin space,
without affecting the lattice point group symmetry), the other two
correspond to simultaneous rotational symmetry breaking in lattice
and in spin space. Clearly, these order parameters mix if one includes
spin-orbit interaction. However, it is still an interesting open question
whether there are iron-based superconductors or other materials where
these quadrupolar order parameters are the dominant vestigial order parameters. 

The above analysis applies to any square-lattice system displaying
density-waves with ordering vectors $\mathbf{Q}_{1}=\left(\pi,0\right)$
and $\mathbf{Q}_{2}=\left(0,\pi\right)$. While we focused on the
case of spin density-waves here, extension to the case of charge density-waves
is straightforward. In particular, in the case of (commensurate) charge
density-waves, the possible vestigial orders are exactly the same
as the $j=0$ composite order parameters discussed above.

\section{VESTIGIAL ORDER FROM DENSITY-WAVES IN THE HEXAGONAL LATTICE}
\label{sec:vestigial-order-from}
A similar analysis as the one outlined above can be performed for
the case of density-waves in the hexagonal lattice. The new aspect
of this problem is the existence of a triply-degenerate ground state,
which allows us to discuss the case where the primary order parameter
transforms as a three-dimensional irreducible representation. In this
situation, non-trivial trilinear composite order parameters can exist,
leading to an even richer phase diagram. We note that the classification
of the primary order parameters for this situation was previously
done in Refs. \cite{Venderbos16,Venderbos2_16}; we follow the notation
of that paper to study the various composite orders.

In the case of spin density-waves, the local spin is parametrized
in terms of three magnetic order parameters $\mathbf{m}_{a}$ associated
with three wave-vectors related by $60^{\circ}$ rotations: $\mathbf{Q}_{1}=\frac{\pi}{\sqrt{3}}\left(\sqrt{3},1\right)$,
$\mathbf{Q}_{2}=\frac{\pi}{\sqrt{3}}\left(0,-2\right)$, and $\mathbf{Q}_{3}=\frac{\pi}{\sqrt{3}}\left(-\sqrt{3},1\right)$,
such that $\mathbf{Q}_{1}+\mathbf{Q}_{2}+\mathbf{Q}_{3}=0$:

\begin{equation}
\mathbf{S}\left(\mathbf{r}\right)=\sum_{a=1,2,3}\mathbf{m}_{a}\cos \bigl( \mathbf{Q}_{a}\cdot\mathbf{r} \bigr)
\end{equation}

\begin{figure}
\begin{centering}
\includegraphics[width=\linewidth]{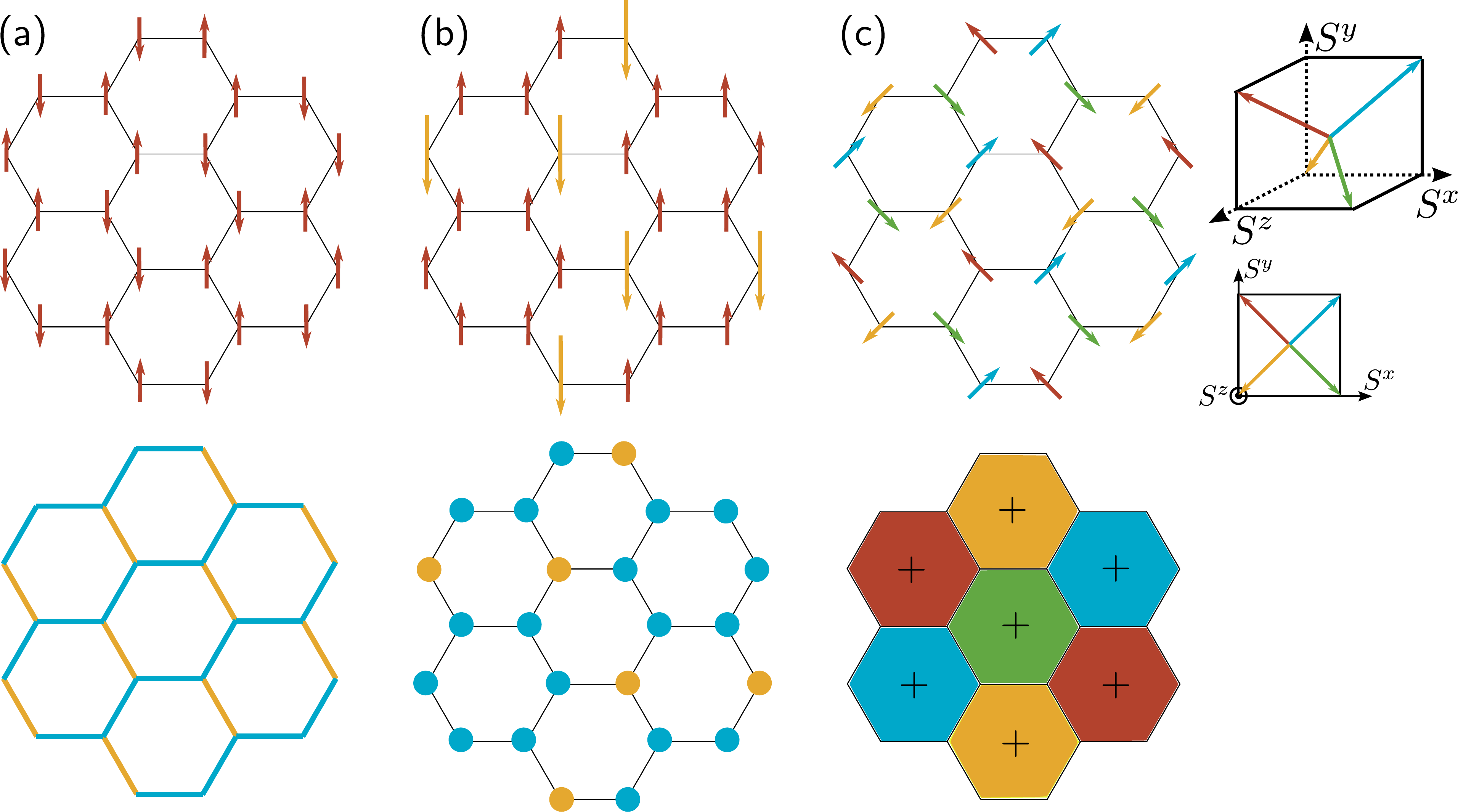}
\par\end{centering}
\caption{Schematic representation of the three possible hexagonal-lattice spin
density-wave ground states (upper panels) with ordering vectors $\mathbf{Q}_{1}=\frac{\pi}{\sqrt{3}}\left(\sqrt{3},1\right)$,
$\mathbf{Q}_{2}=\frac{\pi}{\sqrt{3}}\left(0,-2\right)$, and $\mathbf{Q}_{3}=\frac{\pi}{\sqrt{3}}\left(-\sqrt{3},1\right)$,
and their corresponding vestigial phases (lower panels). Panel (a)
refers to the single-\textbf{Q }magnetic phase and its corresponding
nematic vestigial phase, characterized by unequal bonds. Panel (b)
shows the collinear triple-\textbf{Q }magnetic phase and its corresponding
charge-ordered vestigial phase, characterized by unequal sites. Note that the magnitude of the magnetization is three times larger in some sites (orange arrows) than in other sites (red arrows). Panel
(c) illustrates the non-coplanar triple-\textbf{Q }magnetic phase, 
and its corresponding spin-current vestigial phase. In the magnetically ordered state, the local magnetization can point in one of the four directions shown in the upper inset; for simplicity, here we represent these four directions in the plane, as shown in the lower inset. The vestigial triple-\textbf{Q } spin-current phase is characterized by
unequal plaquettes; specifically, there are four types of plaquettes corresponding to one of the four polarizations of the vector spin-chirality. Note that the scalar spin-chirality $\mathbf{m}_1 \cdot (\mathbf{m}_2 \times \mathbf{m}_3)$ is positive and equal in each plaquette for the spin configuration drawn here. \label{fig_hexagonal_lattice}}
\end{figure}

The three possible magnetic ground states, illustrated in Fig. \ref{fig_hexagonal_lattice},
correspond to \cite{Nandkishore_Chern12}: (i) a single-\textbf{Q}
spin density-wave phase, in which only one of the $\mathbf{m}_{a}$
order parameters is non-zero; (ii) a collinear triple-\textbf{Q} spin
density-wave, in which all three magnetic order parameters are non-zero
and parallel or anti-parallel to each other; (iii) a non-coplanar
triple-\textbf{Q} spin density-wave, in which again all three magnetic
order parameters are non-zero and perpendicular to each other. 

Such a description has been employed to study the magnetic properties
of graphene doped to its van-Hove singularity point and also of doped
cobaltates \cite{TLi12,Nandkishore_Chern12,DHLee12,Chern12,Batista12}.
More recently, it has been applied to SrTiO$_{3}$ thin films grown
along the $(111)$ orientation \cite{Paramekanti18}. It can be derived
from an itinerant microscopic model with a nearly-nested Fermi surface
\cite{Nandkishore_Chern12}.

The relevant group of the primary order parameter is given, in the
absence of spin-orbit coupling, by

\begin{equation}
{\cal G}=C_{6v}^{'''}\times SO\left(3\right).
\end{equation}
The extended point group $C_{6v}^{'''}$ corresponds to the point
group $C_{6v}$ supplemented by three translations $T_{1}=\frac{1}{2}\left(1,\,\sqrt{3}\right)$,
$T_{2}=\frac{1}{2}\left(1,\,-\sqrt{3}\right)$, and $T_{3}=\left(1,\,0\right)$.
The primary order parameters $\mathbf{m}_{a}$ transform according
to the irreducible representation $\Gamma=F_{1}\otimes\Gamma^{S=1}$.
Since $F_{1}$ is a three-dimensional irreducible representation,
we have a nine-dimensional primary order parameter $\eta_{A}=\left(\mathbf{m}_{1},\,\mathbf{m}_{2},\,\mathbf{m}_{3}\right)$.
To proceed and form the bilinear forms, we use Eq. (\ref{eq_Gamma_s})
to decompose the product $\Gamma^{S}\otimes\Gamma^{S}$ and also

\begin{equation}
F_{1}\otimes F_{1}=A_{1}\oplus E_{2}\oplus F_{2}\oplus F_{1}.
\end{equation}

Note that the decomposition of $F_{1}\otimes F_{1}$ does not only
yield one-dimensional (1D) irreducible representations. Instead, in
addition to the trivial irreducible representation $A_{1}$, we obtain
the two-dimensional irreducible representation $E_{2}$ (corresponding
to the degeneracy between $d_{xy}$ and $d_{x^{2}-y^{2}}$ in the
hexagonal lattice), and the three-dimensional irreducible representations
$F_{1}$ and $F_{2}$, which correspond to orders that break translational
symmetry according to the wave-vectors $\mathbf{Q}_{1}$, $\mathbf{Q}_{2}$,
and $\mathbf{Q}_{3}$.

Similarly to the case of the square lattice, we introduce the index
$m=\left(r,j\right)$ that is a combination of the spatial irreducible
representations $r=\left(0,1,2,3\right)=\left(A_{1},E_{2},F_{2},F_{1}\right)$
and the spin index $j$. The bilinears are once again given by

\begin{equation}
\phi_{m\equiv\left(r,j\right)}^{\nu\mu}=\sum_{A,B}\eta_{A}\Lambda_{A,B}^{m,\nu\mu}\eta_{B}
\end{equation}
with $A=\left(a,\alpha\right)$, $B=(b,\beta)$ and matrices: 
\begin{equation}
\Lambda_{A,B}^{m\equiv(r,j),\nu\mu}=\Gamma_{ab}^{r,\nu}\lambda_{\alpha\beta}^{j,\mu}.
\end{equation}

The spin-space matrices $\lambda_{\alpha\beta}^{j,\mu}$ are the same
as the ones presented in the previous section. As for the nine $3\times3$
matrices $\Gamma_{ab}^{r,\nu}$, they can be expressed in terms of
the identity matrix $\Gamma_{ab}^{0,0}=\delta_{ab}$ and the eight
Gell-Mann matrices. Denoting them by the usual notation $\lambda_{ab}^{l}$,
with $l=1,\ldots,8$, we separate the eight matrices into one doublet
and two triplets according to: $\Gamma_{ab}^{1,\nu}=\left\{ \lambda_{ab}^{3},\,\lambda_{ab}^{8}\right\} $,
$\Gamma_{ab}^{2,\nu}=\left\{ \lambda_{ab}^{2},\,\lambda_{ab}^{7},\,\lambda_{ab}^{5}\right\} $,
and $\Gamma_{ab}^{3,\nu}=\left\{ \lambda_{ab}^{1},\,\lambda_{ab}^{6},\,\lambda_{ab}^{4}\right\} $.
In what follows, we focus on scalar and vector bilinears; rank-2 tensor
bilinears can be obtained in the same way as in the previous section
in a straightforward way. 

For $j=0$, the bilinears are scalars and given by $\phi_{(r,0)}^{\nu}=\sum_{a,b=1}^{3}\left(\mathbf{m}_{a}\cdot\mathbf{m}_{b}\right)\Gamma_{ab}^{r,\nu}$.
We find six non-zero possible bilinears:

\begin{align}
\phi_{(0,0)} & =m_{1}^{2}+m_{2}^{2}+m_{3}^{2}\nonumber \\
\phi_{(1,0)}^{\nu} & =\left\{ m_{1}^{2}-m_{2}^{2},\,\frac{1}{\sqrt{3}}\left(m_{1}^{2}+m_{2}^{2}-2m_{3}^{2}\right)\right\} \nonumber \\
\phi_{(3,0)}^{\nu} & =2\left\{ \mathbf{m}_{1}\cdot\mathbf{m}_{2},\,\mathbf{m}_{2}\cdot\mathbf{m}_{3},\,\mathbf{m}_{1}\cdot\mathbf{m}_{3}\right\} \label{eq_phi_hex_DW}
\end{align}

Note that $\phi_{(0,0)}$ transforms trivially as $A_{1}$ and thus
cannot form a vestigial order. The two order parameters of $\phi_{(1,0)}^{\nu}$
transform non-trivially as the two-dimensional irreducible representation
$E_{2}$ and correspond to nematic orders with $d_{x^{2}-y^{2}}$
and $d_{xy}$ form factors, respectively. These bilinears allow for
a vestigial nematic state that lowers the point-group symmetry without
breaking translational symmetry. They are the vestigial phase of the
single-\textbf{Q }spin density-wave (see Fig. \ref{fig_hexagonal_lattice}a). 

The three order parameters of $\phi_{(3,0)}^{\nu}$ transform non-trivially
as the three-dimensional irreducible representation $F_{1}$. They
preserve the point-group symmetry of the lattice but break translational
symmetry. Thus, they correspond to charge density-waves with ordering
vectors $\mathbf{Q}_{3}$ ($\nu=1$), $\mathbf{Q}_{1}$ ($\nu=2$),
and $\mathbf{Q}_{2}$ ($\nu=3$), which are vestigial orders of the
collinear triple-\textbf{Q} spin density-wave (see Fig. \ref{fig_hexagonal_lattice}b).
As shown in Ref.\cite{Chern12}, the transition to the vestigial phase
belongs to the same universality class of the $4$-state Potts model,
corresponding to $\phi_{(3,0)}^{\nu}=\pm1$ subject to the constraint
$\prod\limits _{\nu=1}^{3}\mathrm{sign}\left[\phi_{(3,0)}^{\nu}\right]=\pm1$.
Finally, the composite order parameters $\phi_{(2,0)}^{\nu}$vanish
as the three corresponding Gell-Mann matrices are purely imaginary,
but $m_{a}^{\alpha}$ are real. 

For $j=1$, we obtain vector bilinears according to $\boldsymbol{\phi}_{(r,1)}^{\nu}=\sum_{a,b}\left(\mathbf{m}_{a}\times\mathbf{m}_{b}\right)\Gamma_{ab}^{r,\nu}$.
There are three non-zero such bilinears, which transform as the three-dimensional
irreducible representation $F_{2}$:

\begin{equation}
\boldsymbol{\phi}_{(2,1)}^{\nu}=2\left\{ \mathbf{m}_{1}\times\mathbf{m}_{2},\,\mathbf{m}_{2}\times\mathbf{m}_{3},\,\mathbf{m}_{1}\times\mathbf{m}_{3}\right\} \label{eq_phi_hex_DW2}
\end{equation}

Each of them corresponds to spin-current density-waves (i.e. vector
chirality) with ordering vectors $\mathbf{Q}_{3}$ ($\nu=1$), $\mathbf{Q}_{1}$
($\nu=2$), and $\mathbf{Q}_{2}$ ($\nu=3$). The resulting vestigial order is thus the triple-\textbf{Q} spin-current order shown in Fig. \ref{fig_hexagonal_lattice}c.
This is the vestigial phase of the non-coplanar triple-\textbf{Q} spin density-wave.

Interestingly, because the primary order parameter transforms as
a three-dimensional irreducible representation, it is possible to
also construct trilinear forms $\psi^{m}=\sum_{A,B,C}\eta_{A}\eta_{B}\eta_{C}\Lambda_{A,B,C}^{m}$
that transform non-trivially. This can be formally done by combining
vestigial order parameters $\phi_{\left(r,j\right)}^{\nu\mu}$ that
transform as higher-dimensional irreducible representations and the
primary order parameter. Among the bilinears presented in Eqs. (\ref{eq_phi_hex_DW})
and (\ref{eq_phi_hex_DW2}), combining $\boldsymbol{\phi}_{(2,1)}^{\nu}$,
which transforms as $F_{2}$, with the primary order parameter $\eta_{A}$,
which transforms as $F_{1}$, yields a composite trilinear scalar
that transforms non-trivially according to the $A_{2}$ irreducible
representation, since $F_{1}\otimes F_{2}=E_{2}\oplus F_{1}\oplus F_{2}\oplus A_{2}$.
In explicit form, the corresponding order parameter $\psi$ is given
by:

\begin{equation}
\psi=\mathbf{m}_{1}\cdot\left(\mathbf{m}_{2}\times\mathbf{m}_{3}\right).
\end{equation}

We identify $\psi$ as the scalar chirality, an Ising-like, $\mathbf{Q}_{1}+\mathbf{Q}_{2}+\mathbf{Q}_{3}=0$,
order parameter that breaks time-reversal symmetry \cite{Martin08,Venderbos2_16}.
Similarly to the bilinear $\boldsymbol{\phi}_{(2,1)}^{\nu}$, it is
also a vestigial phase of the non-coplanar triple-\textbf{Q }spin
density-wave. This brings an interesting scenario, in which there
are two different vestigial phases associated with the same primary
order. While $\psi$ breaks a discrete symmetry, the vector chirality
$\boldsymbol{\phi}_{(2,1)}^{\nu}$ is a continuous order parameter.
One thus expects the vestigial scalar chirality $\psi$ to order at
a higher temperature than the vestigial spin-current density-waves
$\boldsymbol{\phi}_{(2,1)}^{\nu}$ in a sufficiently strongly anisotropic
three-dimensional system (as schematically shown in Fig. \ref{fig_phase_diagrams}e).
A microscopic calculation of such a scenario remains to be seen.

We finish this section by discussing the case in which the primary
order parameter is a charge density-wave. In this situation, one would
expect vestigial orders corresponding to the $j=0$ composite order
parameters of the spin density-wave case, namely, $\phi_{(1,0)}^{\nu}$
and $\phi_{(3,0)}^{\nu}$ in Eq. (\ref{eq_phi_hex_DW}). However,
$\phi_{(3,0)}^{\nu}$ corresponds to charge density-waves with the
same ordering vectors as the primary order parameters, and thus do
not constitute a vestigial order. Moreover, $\phi_{(1,0)}^{\nu}$
cannot be realized, since it is not possible to form a single-\textbf{Q
}charge-density wave. This follows from the fact that the trilinear
$W_{1}W_{2}W_{3}$ (where $W_{a}$ correspond to the Ising-like charge
density-wave order parameters) transforms trivially as $A_{1}$, implying
that only triple-\textbf{Q }charge density-waves can be formed in
the hexagonal lattice. As a result, even though the ground state is
degenerate, in this particular case no vestigial order appears. Similar
arguments imply that the nematic phase alone, which transforms as
the two-dimensional irreducible representation $E_{2}$, Eq. (\ref{eq_phi_hex_DW}),
does not admit vestigial phases. This is because a cubic term appears
in the free energy selecting the $d_{x^{2}-y^{2}}$ over the $d_{xy}$
nematic state \cite{Hecker2018}.

\section{OTHER EXAMPLES OF VESTIGIAL ORDER}
\label{sec:other-exampl-vest}
Besides the examples discussed above, there are several other systems
that allow vestigial orders to appear. Here we discuss some of them,
without the same level of details as in the previous sections. The
example that we analyzed for the square lattice consisted of doubly-degenerate
spin density-wave ordering vectors $\left(\pi,0\right)$ and $\left(0,\pi\right)$.
We mentioned that a non-degenerate ground state, such as the N\'eel
order, which displays ordering vector $\left(\pi,\pi\right)$, does
not allow vestigial orders that lower the point-group symmetry of
the lattice. Yet, this does not imply that vestigial order is impossible
for primary N\'eel order. Similarly to the composite order with $j=2$
discussed in Eq. (\ref{eq_tensor_rank2}), in the case of N\'eel order
it is possible to form a rank-2 tensor bilinear analogous to $\phi_{\left(0,2\right)}^{\mu\mu'}$
that breaks spin-rotational invariance while preserving the point
group. This so-called spin-nematic phase \cite{Andreev84} has been
widely discussed in the context of spin-1 models. A candidate material
for spin-nematic order is NiGa$_{2}$S$_{4}$\cite{Nakatsuji2005},
which can be described by spin-1 Heisenberg spins on a triangular
lattice. 

\begin{figure}
\begin{centering}
\includegraphics[width=\linewidth]{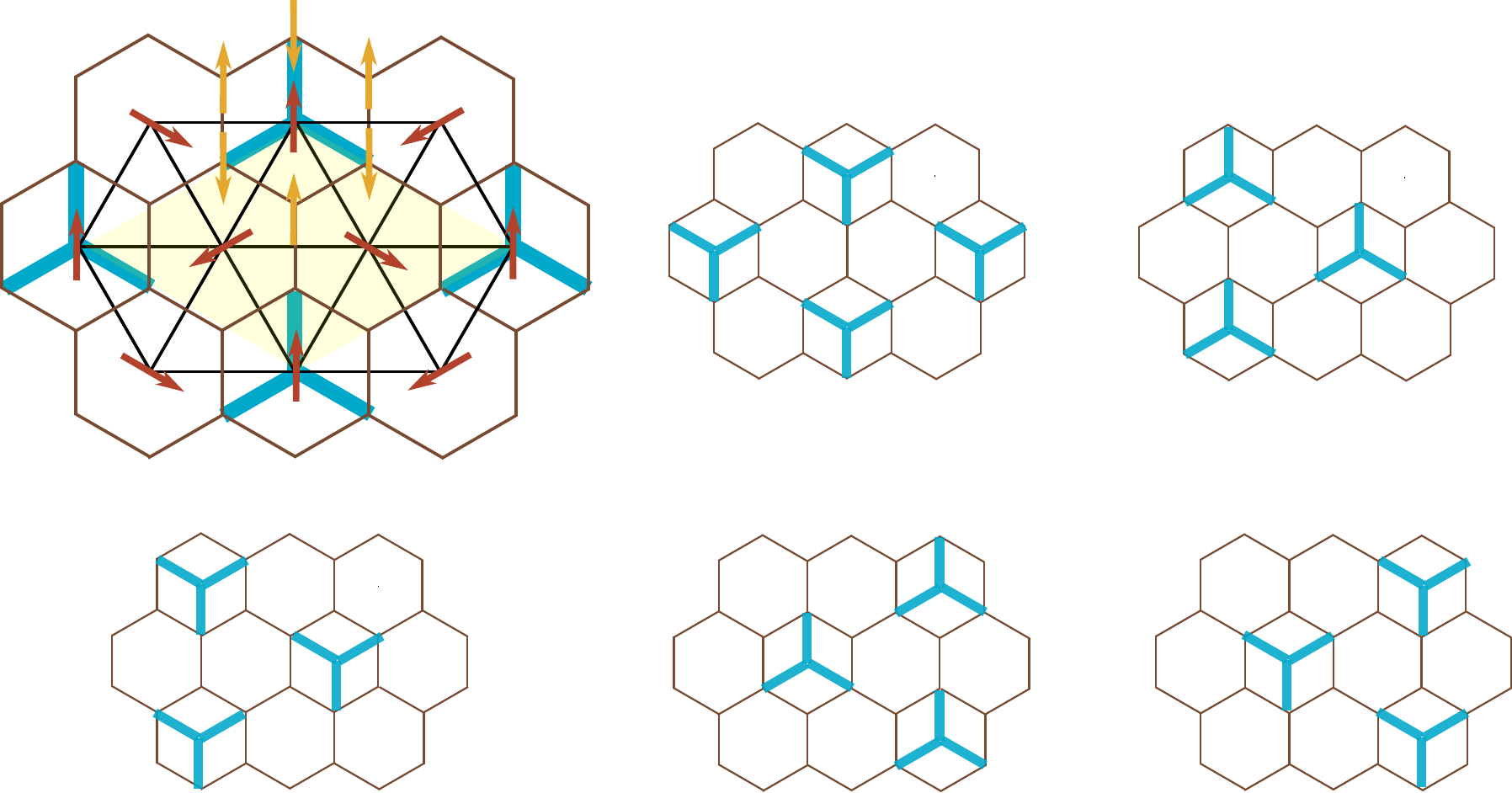}
\par\end{centering}
\caption{Schematic representation of the magnetic ground state of the $J_{1}$-$J_{2}$
model on the windmill lattice, which is composed of interpenetrating
triangular (black lines) and honeycomb (dark brown lines) sublattices.
The spins order in a N\'eel pattern in the honeycomb sublattice and
in a $120^{\circ}$ coplanar configuration in the triangular sublattice.
The vestigial order is described in terms of a $\mathbb{Z}_{6}$ clock-model
order parameter, and corresponds to different relative orientations
between the spin orders of the two sublattices (blue bonds denote antiparallel spins). \label{fig_windmill}}

\end{figure}

Still focusing on the square lattice, it is also possible to have
magnetic ground states with degeneracy higher than $2$, such as the
fourfold-degenerate ordering vectors $\left(\frac{\pi}{2},\frac{\pi}{2}\right)$,
$\left(-\frac{\pi}{2},\frac{\pi}{2}\right)$, $\left(\frac{\pi}{2},-\frac{\pi}{2}\right)$,
and $\left(-\frac{\pi}{2},-\frac{\pi}{2}\right)$. Microscopically,
such states arise for instance in the $J_{1}$-$J_{2}$-$J_{3}$ Heisenberg
model on the square lattice, with dominant $J_{3}$ . Besides the
so-called double-stripe magnetic ground states, plaquette ground states
are also realized in this model \cite{Ducatman12}. It was shown in
Ref. \cite{Zhang17} that the double-stripe magnetic state has two
Ising-like vestigial orders, corresponding to a $B_{2g}$ nematic
order and a $\left(\pi,\pi\right)$ bond-order that breaks translational
and reflection symmetries of the lattice. Experimentally, material
candidates to exhibit these vestigial orders are the iron chalcogenide
FeTe~\cite{FeTe1,FeTe2,JPHu09,Paul11} and the titanium-based
oxypnictide BaTi$_{2}$Sb$_{2}$O \cite{Zhang17}. A full classification
of all possible vestigial orders in this case is not yet available.

Another interesting case of multiple vestigial orders is that of the
$J_{1}$-$J_{2}$ Heisenberg antiferromagnet on the windmill lattice,
which consists of interpenetrating triangular and honeycomb sublattices
\cite{Orth12,Orth14,Jeevanesan15}. The $J_{1}$-$J_{2}$ windmill
model is a straightforward, but non-trivial generalization of the
corresponding $J_{1}$-$J_{2}$ square lattice model. It hosts a vestigial
$Z_{6}$ clock order parameter instead of the Ising nematic $Z_{2}$
degree of freedom found on the square lattice. In both cases, vestigial
order refers to a \textit{\emph{relative}} ordering of spins on the
two sublattices. In the windmill case, it breaks translation symmetry
by tripling the unit cell as well as a mirror symmetry that exchanges
the $A$ and $B$ sites on the honeycomb sublattice\textbf{ (}see
Fig. \ref{fig_windmill}). Interestingly, due to the higher degree
of degeneracy of the composite order parameter ($Z_{n}$ with $n\geq5$),
vestigial long-range order develops via a two-step process consisting
of two Kosterlitz-Thouless phase transitions that enclose an intermediate
critical phase. In the critical phase, the correlations of the composite
degrees of freedom decay algebraically with a temperature-dependent
exponent $\eta(T)$\cite{Jeevanesan15} \textbf{(}see Fig. \ref{fig_phase_diagrams}f).
Such a behavior is reminiscent of melting of two-dimensional solids,
where (algebraic) translational order disappears via an intermediate
hexatic phase\cite{Halperin78}.

In all the density-wave examples discussed so far, only commensurate
ordering vectors were considered. In the case of incommensurate wave-vectors,
the primary order parameters become complex, which can lead to vestigial
orders that break time-reversal symmetry. This was proposed for instance
in Ref. \cite{Chubukov14} in the context of charge density-waves
in the cuprates. Moreover, incommensurate spin density-wave with wave-vector
$\mathbf{Q}$ naturally couples to incommensurate charge density-wave
with wave-vector $2\mathbf{Q}$, intrinsically coupling charge-driven
and spin-driven nematicity. A full microscopic description of this
peculiar case, which may be relevant for the cuprate La$_{2-x}$Sr$_{x}$CuO$_{4}$,
is still missing \cite{Nie17}. The case of incommensurate charge
density-wave also highlights the key role of disorder in vestigial
phases: as shown in Ref. \cite{Nie14}, in tw -dimensions, any amount
of disorder kills incommensurate charge order at finite temperatures,
while preserving its nematic vestigial phase. Overall, the impact
of disorder on vestigial phases remains little explored, despite the
ubiquity of disorder in realistic systems (see also Ref. \cite{Cui18}).

It is also important to emphasize that the existence of degenerate
ground states is a necessary, but not sufficient condition for the
appearance of vestigial orders. We already mentioned this feature
in the case of charge density-waves in the hexagonal lattice, which
do not allow any vestigial phases. Another interesting example is
the case of spin-dimers on a square lattice with nearest and next-nearest
neighbor interactions and in the presence of an external magnetic
field \cite{Loison2000}. A closely related model has been employed
to describe the unusual properties of BaCuSi$_{2}$O$_{6}$ in an
external magnetic field \cite{Sebastian06,Batista07,Schmalian08}.
The model is equivalent to hard-core bosons on a square lattice that
can undergo Bose-Einstein condensation at momenta $\mathbf{Q}_{1}=\left(\pi,0\right)$
or $\mathbf{Q}_{2}=\left(0,\pi\right)$. As the chemical potential
$\mu$ of these hard-core bosons is tuned from negative to positive,
there is a quantum phase transition from a disordered state to a condensate
with finite momentum. The model thus shares the same properties as
the case of doubly-degenerate charge density-waves on the square lattice
discussed above. However, there is one important difference: at $T=0$
and at the quantum critical point ($\mu=0$), the system is empty
of bosons and fluctuations are thus irrelevant. As a result, there
are no fluctuations to trigger a $T=0$ vestigial phase. Formally,
this is manifested in the Ginzburg-Landau free-energy expansion by
the vanishing of all the coefficients in front of the squared non-trivial
bilinears (see for instance Eq. \eqref{SC_free_energy1}). The resulting
phase diagram has therefore a vestigial phase of the finite-momentum
Bose-Einstein condensate only at finite temperatures, but a single
second-order phase transition at $T=0$ (see Fig. \ref{fig_phase_diagrams}c). 

\section{CONCLUDING REMARKS}
\label{sec:concluding-remarks}

The formalism developed here demonstrates that multi-component order
parameters give rise to complex phase diagrams, providing an appealing
framework to understand quantum materials that goes beyond the paradigm
of competing phases. The degenerate nature of the ordered state \textendash{}
a necessary but not sufficient condition for the emergence of vestigial
order \textendash{} leads to the condensation of fluctuations at their
own transition temperature, manifested by long-range order of composite
operators. In this regard, composide order not only behaves as a vestige
of the primary order, but it also affects the latter by lifting its
degeneracy and thus relieving the frustration of the system. In situations
where the primary order cannot establish long-range order, either
due to strong thermal or quantum fluctuations or due to disorder,
the vestigial order is the only sharp remanent of the primary order. 

While symmetry arguments can efficiently be employed to classify which
vestigial states are allowed in each case, they do not prove the actual
existence of vestigial phases. Only via microscopic calculations of
minimal models one can assess whether the vestigial and primary phase
transitions take place at different temperatures or simultaneously
as a first-order transition \textendash{} in which case there is no
vestigial order. Particularly near a quantum phase transition, the
symmetry of the order parameter is not enough to determine the final
behavior of the vestigial phase, as the dynamics of the primary order
parameter plays an essential role. Theoretically, while mean-field
calculations are incapable of capturing vestigial phases, a variety
of controlled and uncontrolled analytical methods exist, such as the
saddle-point large-$N$ approach \cite{Fernandes2012,Nie14}, the
self-consistent Gaussian approximation \cite{Fischer16,Nie17}, and
the renormalization-group approach \cite{Qi09,Millis10}. Numerically,
vestigial order can be addressed straightforwardly by analyzing the
statistical properties of the corresponding higher-order correlation
functions. For example, the Ising nematic transition in the classical
$J_{1}$-$J_{2}$ square lattice model has been directly observed
using Monte-Carlo simulations \cite{Weber03,Batista11} (see Ref.\cite{Jeevanesan15}
for a related study on the windmill lattice). An interesting further
direction is to investigate vestigial order in low-dimensional quantum
(spin) systems at zero temperature, where powerful numerical techniques
are available\cite{Stoudenmire12}. 

The concept of vestigial order has thus the potential to be applied
to a vast number of systems that have been partially explored or that
even remain completely unexplored. An interesting issue that goes
beyond broken-symmetry phases is whether topologically-driven orders
may also support unusual vestigial states of matter \cite{Scheurer17}.




\section*{ACKNOWLEDGMENTS}
We thank C. Batista, E. Berg, P. Chandra, G. W. Chern, A. Chubukov,
M. Christensen, P. Coleman, I. Eremin, R. Flint, E. Fradkin, M. Hecker,
B. Jeevanesan, J. Kang, S. Kivelson, I. Mazin, J. Venderbos, and X.
Wang for fruitful discussions and collaborations on topics reviewed
in this work. R.M.F. is supported by the US Department of Energy, Office
of Science, Basic Energy Sciences, under Award DE-SC0012336. P.P.O. acknowledges
support from Iowa State University Startup Funds. J.S. is supported
by the Helmholtz Program \emph{Science and Technology of Nanosystems}
at the Karlsruhe Institute of Technology (KIT). 


\end{document}